\documentclass[aps,prb,twocolumn,showpacs,byrevtex]{revtex4}

\usepackage{times,mathptm}
\usepackage[dvips]{graphicx,color}

\begin{document}
\title{Spin-orbit coupling effects on the stability of two competing structures in Pb/Si(111) and Pb/Ge(111)}
\author{Xiao-Yan Ren$^{1,2}$, Hyun-Jung Kim$^3$, Seho Yi$^2$, Yu Jia$^{1,4*}$, and Jun-Hyung Cho$^{2,1*}$}
\affiliation{$^1$ International Laboratory for Quantum Functional Materials of Henan, and School of Physics and Engineering, Zhengzhou University, Zhengzhou 450001, China \\
$^2$ Department of Physics and Research Institute for Natural Sciences, Hanyang University,
17 Haengdang-Dong, Seongdong-Ku, Seoul 133-791, Korea \\
$^3$ Korea Institute for Advanced Study, 85 Hoegiro, Dongdaemun-gu, Seoul 130-722, Korea \\
$^4$ Key Laboratory for Special Functional Materials of Ministry of Education, School of Physics and Electronics, Henan University, Kaifeng 475004, China}
\date{\today}

\begin{abstract}
Using first-principles density-functional theory (DFT) calculations, we investigate the 4/3-monolayer structure of Pb on the Si(111) or Ge(111) surface within the two competing structural models termed the H$_3$ and T$_4$ structures. We find that the spin-orbit coupling (SOC) influences the relative stability of the two structures in both the Pb/Si(111) and Pb/Ge(111) systems: i.e., our DFT calculation without including the SOC predicts that the T$_4$  structure is energetically favored over the H$_3$ structure by ${\Delta}E$ = 25 meV for Pb/Si(111) and 22 meV for Pb/Ge(111), but the inclusion of SOC reverses their relative stability as ${\Delta}E$ = $-$12 and $-$7 meV, respectively. Our analysis shows that the SOC-induced switching of the ground state is attributed to a more asymmetric surface charge distribution in the H$_3$ structure, which gives rise to a relatively larger Rashba spin splitting of surface states as well as a relatively larger pseudo-gap opening compared to the T$_4$ structure. By the nudged elastic band calculation, we obtain a sizable energy barrier from the H$_3$ to the T$_4$ structure as ${\sim}$0.59 and ${\sim}$0.27 eV for Pb/Si(111) and Pb/Ge(111), respectively. It is thus likely that the two energetically competing structures can coexist at low temperatures.
\end{abstract}


\maketitle

\section{INTRODUCTION}

Two-dimensional (2D) electronic systems have attracted intensive attention due to their many exotic phenomena and potential applications in nanoelectronic devices.~\cite{brun,zhang,kurosaki,car,coldea,wei,meng,shimizu} Recently, metal-atom adsorption on semiconductor surfaces has been employed to generate spin-polarized current on the basis of the Rashba spin splitting, where the spin-orbit coupling (SOC) lifts the spin degeneracy due to the broken inversion symmetry.~\cite{kim,matet,barke,gierz,frant,stolw,gluz,stol} We here focus on the 4/3 monolayer (ML) adsorption of Pb atoms on the Si(111) or Ge(111) surface, forming a dense phase with the ${\sqrt{3}}{\times}{\sqrt{3}}R30^{\circ}$ unit cell [see Figs. 1(a) and 1(b)]. It was recently reported~\cite{yaji} that this phase displays a large spin splitting of metallic surface bands on the semiconducting Ge(111) substrate, thereby promising for surface spin transport/accumulation because surface spin signals can be preserved over long time within the substrate band gap.

Earlier reflection high energy electron diffraction (RHEED), x-ray, and low energy electron diffraction (LEED) measurements for Pb/Ge(111) suggested that the 4/3 ML coverage has four Pb atoms per unit cell, three of which is located at off-centered T$_1$ sites and the fourth Pb atom at an H$_3$ site.~\cite{dev,feid,huang} This structural model is termed the H$_3$ structure [see Fig. 1(a)], which preserves the $C_{3v}$ symmetry. Recently, angle-resolved photoelectron spectroscopy (ARPES) and spin-resolved ARPES for Pb/Ge(111) demonstrated that a metallic surface-state band with a dominant Pb 6$p$ character exhibits a large Rashba spin splitting of 200 meV.~\cite{yaji} This observed spin splitting of surface state agreed well with the surface band structure of the H$_3$ structure (equivalently the close-packed structure), obtained using a density-functional theory (DFT) calculation with including the SOC~\cite{yaji,yaji2}. However, earlier DFT calculations without SOC for Pb/Si(111)~\cite{chan,hupa} and Pb/Ge(111)~\cite{anci} predicted that the T$_4$ structure [see Fig. 1(b)] or the so-called chain structure with the broken $C_{3v}$ symmetry is slightly more stable than the H$_3$ structure by less than a few tens of meV. This conflict of the ground state between experiments~\cite{dev,feid,huang} and previous DFT calculations~\cite{chan,hupa,anci} may imply that the SOC influences the relative stability of the H$_3$ and T$_4$ structures in the Pb/Si(111) and Pb/Ge(111) surface systems. Therefore, it is very challenging to examine the energetics of the two structures by using the DFT calculations with/without including the SOC.

\begin{figure}[ht]
\centering{ \includegraphics[width=8.0cm]{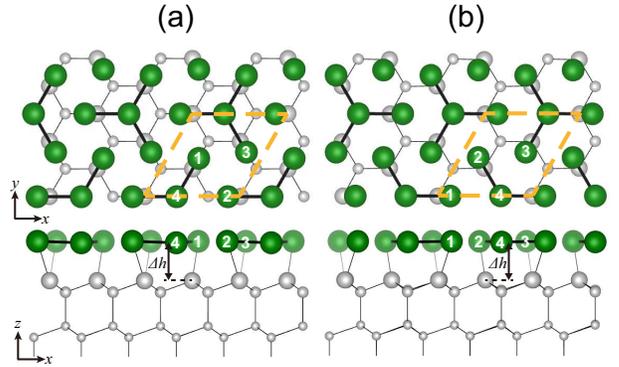} }
\caption{(Color online) Top and side views of the optimized (a) H$_3$ and (b) T$_4$ structures of Pb/Si(111). The green and gray circles represent Pb and Si atoms, respectively. For distinction, the Si atoms in the subsurface layers are drawn with small circles. The dashed line indicates the ${\sqrt{3}}{\times}{\sqrt{3}}R30^{\circ}$ unit cell, where the four Pb atoms are numbered. The Pb$_4$ atom in (a) and (b) is located at the H$_3$ and T$_4$ sites, respectively. The {\bf x}, {\bf y}, and {\bf z} axes point along the [11${\overline{2}}$], [1${\overline{1}}$0], and [111] directions, respectively. The height difference between the H$_3$ (T$_4$) Pb atom and the first Si-substrate layer is labeled as ${\Delta}h$. }

\end{figure}

In the present study, we optimize the geometries of Pb/Si(111) and Pb/Ge(111) within the H$_3$ and T$_4$ structural models. Our DFT calculation without SOC demonstrates that the T$_4$ structure is energetically favored over the H$_3$ structure by ${\Delta}E$ = 25 meV for Pb/Si(111) and 22 meV for Pb/Ge(111), consistent with previous DFT calculations.~\cite{chan,hupa,anci} However, the inclusion of SOC reverses their relative stability as ${\Delta}E$ = $-$12 and $-$7 meV for Pb/Si(111) and Pb/Ge(111), respectively. We find that the H$_3$ structure has a more asymmetric surface charge distribution compared to the T$_4$ structure, which in turn results in a relatively larger Rashba spin splitting as well as a relatively larger pseudo-gap opening along the $\overline{KM}$ symmetry line. These results provide an explanation for the SOC-driven switching of the ground state. Moreover, our nudged elastic band calculation obtains an energy barrier for the transition on going from the H$_3$ to the T$_4$ structure as large as ${\sim}$0.59 and ${\sim}$0.27 eV for Pb/Si(111) and Pb/Ge(111), respectively. This suggests that the two energetically competing structures can coexist at low temperatures.

\section{CALCULATIONAL METHODS}

The present DFT calculations were performed using the Vienna {\it ab initio} simulation package with the projector-augmented wave method and a plane wave basis set.~\cite{vasp1,vasp2} For the treatment of exchange-correlation energy, we employed the GGA functional of Perdew-Burke-Ernzerhof.~\cite{pbe} The Pb/Si(111) and Pb/Ge(111) surfaces were modeled by a periodic slab geometry consisting of the twelve Si and Ge atomic layers with ${\sim}$25 {\AA} of vacuum in between the slabs. The bottom of the Si or Ge substrate was passivated by one H atom. We employed a dipole correction that cancels the artificial electric field across the slab.~\cite{neugebauer} The kinetic energy cutoff of the plane-wave basis set was taken to be 400 eV, and the ${\bf k}$-space integration was done with the 15${\times}$15 Monkhorst-Pack meshes in the surface Brillouin zones (SBZ) of the ${\sqrt{3}}{\times}{\sqrt{3}}R30^{\circ}$ unit cell. All atoms except the bottom two substrate layers were allowed to relax along the calculated forces until all the residual force components were less than 0.01 eV/{\AA}.

\section{RESULTS}

We begin to optimize the H$_3$ and T$_4$ structures of the Pb/Si(111) and Pb/Ge(111) surfaces using the DFT calculation in the absence of SOC. The optimized H$_3$ and T$_4$ structures of Pb/Si(111) are displayed in Fig. 1(a) and 1(b), respectively. We find that the T$_4$ structure is energetically more stable than the H$_3$ structure by ${\Delta}E$ = 25 and 22 meV per ${\sqrt{3}}{\times}{\sqrt{3}}R30^{\circ}$ unit cell for Pb/Si(111) and Pb/Ge(111), respectively. This preference of the T$_4$ structure over the H$_3$ structure is consistent with previous DFT calculations~\cite{chan,hupa,anci}. The calculated bond lengths $d_{\rm Pb-Pb}$ and $d_{\rm Pb-Si}$ (or $d_{\rm Pb-Ge}$) at the interface are given in Table I. It is noticeable that the H$_3$ structure of Pb/Si(111) [Pb/Ge(111)] has an equal value of $d_{\rm Pb-Pb}$ = 3.137 (3.224) {\AA} between three nearest Pb$-$Pb atoms, thereby preserving the $C_{3v}$ symmetry. Meanwhile, in the T$_4$ structure of Pb/Si(111) [Pb/Ge(111)], $d_{\rm Pb-Pb}$ = 3.183 (3.265) {\AA} between two nearest Pb$-$Pb atoms is slightly shorter than the third one by ${\sim}$0.01 (0.02) {\AA}, indicating the broken $C_{3v}$ symmetry. This rotational symmetry breaking in the T$_4$  structure is also reflected by the position of the T$_4$ Pb adatom which is located at the off-centered T$_4$ site with a lateral displacement of ${\sim}$0.05 {\AA} along the {\bf x} direction [see Fig. 1(b)].

\begin{table}[ht]
\caption{Bond lengths (in {\AA}) $d_{\rm Pb-Pb}$ and $d_{\rm Pb-Si}$ (or $d_{\rm Pb-Ge}$) as well as ${\Delta}h$ (see Fig. 1) in the H$_3$ and T$_4$ structures, obtained using DFT and DFT+SOC. Two different values of $d_{\rm Pb-Pb}$ and $d^{'}_{\rm Pb-Pb}$ are given in the T$_4$ structure. The results for Pb/Ge(111) are given in parentheses.}
\begin{ruledtabular}
\begin{tabular}{cccc}
              &  	     & DFT 	    & DFT+SOC   	\\  \hline
   H$_3$              & $d_{\rm Pb-Pb}$  &3.137 (3.224)    & 3.101 (3.188)	  \\
   	         & d$_{{\rm Pb_1}-{\rm Si}}$     &2.888 (2.893)    & 2.882 (2.897)	  \\
   	         & d$_{{\rm Pb_2}-{\rm Si}}$     &2.888 (2.893)    & 2.882 (2.897)	  \\
           	 & d$_{{\rm Pb_3}-{\rm Si}}$     &2.888 (2.893)    & 2.882 (2.897)     \\
                    & ${\Delta}h$         &2.596 (2.737)  & 2.649 (2.761)	  \\   \hline
 T$_4$           & $d_{\rm Pb-Pb}$   & 3.183 (3.265)    & 3.133 (3.218)	  \\
            & $d^{'}_{\rm Pb-Pb}$   & 3.193 (3.281)    &  3.142 (3.237)	  \\
       	     & d$_{{\rm Pb_1}-{\rm Si}}$      &2.877 (2.893)    & 2.867 (2.887)	  \\
  	         & d$_{{\rm Pb_2}-{\rm Si}}$      &2.882 (2.914)    & 2.871 (2.901)	  \\
	         & d$_{{\rm Pb_3}-{\rm Si}}$      &2.882 (2.914)    & 2.871 (2.901)      \\
             & ${\Delta}h$        &2.534 (2.537)    & 2.689 (2.672)	  \\  \hline
\end{tabular}
\end{ruledtabular}
\end{table}

Figures 2(a) and 2(b) show the band structures of the H$_3$ and T$_4$ structural models of Pb/Si(111), respectively, obtained using the DFT calculation without SOC: For Pb/Ge(111), see Figs. 1S(a) and 1S(b) of the Supplemental Material.~\cite{supp} The band projection onto the Pb $p_x$, $p_y$, and $p_z$ orbitals is also displayed in Figs. 2(a) and 2(b). It is seen that the band structures of the H$_3$ and T$_4$ structures are very similar to each other. Especially, the band dispersion of the surface states crossing the Fermi level $E_F$ is nearly parabolic along the $\overline{{\Gamma}M}$ and $\overline{{\Gamma}K}$ lines, indicating 2D electronic states in the Pb overlayer. As shown in Figs. 2(a) and 2(b), the orbital character of the surface states exhibits a strong ${\bf k}$-dependence along the symmetry lines. Consequently, the SOC may induce an efficient hybridization of the Pb $p_x$, $p_y$, and $p_z$ orbitals, as discussed below.

\begin{figure*}[ht]
\centering{ \includegraphics[width=17.0cm]{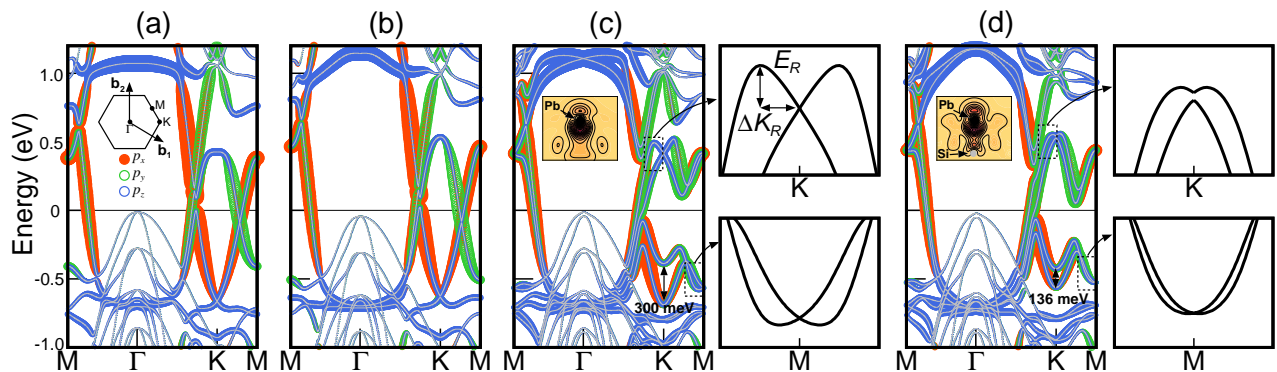} }
\caption{(Color online) Band structures of the (a) H$_3$ and (b) T$_4$ structural models, obtained using DFT. The corresponding ones obtained using DFT+SOC are given in (c) and (d), respectively. The bands projected onto the Pb $p_x$, $p_y$, and $p_z$ orbitals are displayed with circles whose radii are proportional to the weights of such orbitals. The energy zero represents $E_F$. The inset in (a) shows the surface Brillouin zones of the ${\sqrt{3}}{\times}{\sqrt{3}}R30^{\circ}$ unit cell. In (c) and (d), the contour plots of the charge density of the lowest unoccupied surface state at the $K$ point are drawn in the vertical plane containing the H$_3$ and T$_4$ Pb atoms, respectively, where the first line is at 0.3${\times}$10$^{-3}$ electrons/{\AA}$^3$ with spacings of 0.5${\times}$10$^{-3}$ electrons/{\AA}$^3$. A close up of the spin splitting of the surface state at the $K$ point (just above $E_F$) and the $M$ point (just below $E_F$) is displayed for the H$_3$ and T$_4$ structures in (c) and (d), respectively, together with including their time-reversal counterparts.}
\end{figure*}

Next, we examine the SOC effects on the stability of the H$_3$ and T$_4$ structures using the DFT+SOC calculation. Interestingly, the inclusion of SOC changes the relative stability of the two structures: i.e., the H$_3$ structure becomes more stable than the T$_4$ structure by 12 and 7 meV for Pb/Si(111) and Pb/Ge(111), respectively. As shown in Table I, for the H$_3$ (T$_4$) structure of Pb/Si(111), the SOC decreases $d_{\rm Pb-Pb}$ and $d_{\rm Pb-Si}$ by ${\sim}$0.03 (0.05) and ${\sim}$0.01 (0.01) {\AA}, respectively, but increases ${\Delta}h$ by ${\sim}$0.05 (0.16) {\AA}. On the other hand, for the H$_3$ (T$_4$) structure of Pb/Ge(111), the SOC produces decreases of ${\sim}$0.04 (0.05) and ${\sim}$0.01 (0.01) {\AA} for $d_{\rm Pb-Pb}$ and $d_{\rm Pb-Si}$, respectively, and an increase of ${\sim}$0.03 (0.14) {\AA} for ${\Delta}h$. To understand the SOC-induced switching of the ground state, we plot the band structures of the H$_3$ and T$_4$ models of Pb/Si(111) in Fig. 2(c) and 2(d), respectively: For Pb/Ge(111), see Figs. 1S(c) and 1S(d) of the Supplemental Material.~\cite{supp} The spin degeneracy of surface states as well as other states is found to be lifted over the SBZ except at the high-symmetry points (i.e., ${\Gamma}$ and $M$ points). Specifically, it is seen that the H$_3$ structure has a spin splitting of 300 meV for the surface states at the $K$ point just below $E_F$, larger than the corresponding one (136 meV) of the T$_4$ structure. The insets of Figs. 2(c) and 2(d) show a close up of the spin splitting of the surface state at the $K$ point along the $\overline{K{\Gamma}}$ direction (just above $E_F$) and the $M$ point along the $\overline{MK}$ direction (just below $E_F$), obtained for the H$_3$ and T$_4$ structures, respectively. Obviously, these spin-split subbands illustrate the typical dispersion of the Rashba-type spin splitting. We fit the $k$-dependent dispersion of such spin-split subbands with the characteristic parameters such as the momentum offset ${\Delta}k_R$ and the Rashba energy $E_R$ [see Fig. 2(c)] by using the Rashba spin-splitting eigenvalues ${\epsilon}_{\pm}$ = $\frac{{{\hbar}^2}k^2}{2m^*} {\pm} {{\alpha}_R}k$, where $m^*$ is the electron effective mass and ${\alpha}_R$ the Rashba parameter. We find that the H$_3$ structure has ${\alpha}_R$ = 2.775 (0.495) eV {\AA} along $\overline{K{\Gamma}}$ ($\overline{MK}$), larger than 0.523 (0.093) eV {\AA} for the T$_4$ structure. It is also noticeable that the H$_3$ structure exhibits a relatively larger pseudo-gap opening of 385 meV along the $\overline{KM}$ line compared to that (275 meV) of the T$_4$ structure: see Figs. 2(c) and 2(d). Thus, these relatively larger spin splitting and pseudo-gap opening of the H$_3$ structure compared to the T$_4$ structure can lead to the SOC-driven switching of the ground state.

From the DFT+SOC calculation, the band projection of surface states shows a strong hybridization between the Pb $p_x$, $p_y$, and $p_z$ orbitals [see Fig. 2(c) and 2(d)]. This SOC-induced hybridization is known to give not only a gap opening~\cite{sunwoo} but also an asymmetric surface charge distribution~\cite{kim,nagan} which in turn determines the size of the Rashba spin splitting through the integral of the charge density times the potential gradient along the direction perpendicular to the surface, ${\alpha}_R$ ${\propto}$ ${\int} dV/dz\,{\rho}({\bf{r}}) d{\bf{r}}$. As shown in the inset of Figs. 2(c) and 2(d), the H$_3$ structure has a more asymmetric charge character of the surface state compared to the T$_4$ structure, thereby giving rise to the above-mentioned relatively larger values of ${\alpha}_R$. Indeed, a recent DFT calculation for the H$_3$ structure of Pb/Ge(111) reported that the observed large Rashba spin splitting is due to an asymmetric charge distribution in the vicinity of the H$_3$ Pb atom.~\cite{yaji2} It is noteworthy that the SOC changes more dominantly the position of the T$_4$ Pb atom, which moves outward from the surface by ${\sim}$0.16 and ${\sim}$0.14 {\AA} (see Table I) for Pb/Si(111) and Pb/Ge(111), respectively. This outward movement possibly contributes to the rather symmetric charge character of the surface state in the vicinity of the T$_4$ Pb atom, as shown in the inset of Fig. 2(d).

Figures 3(a) and 3(b) show the helical spin textures of the H$_3$ and T$_4$ structures of Pb/Si(111) along the Fermi surface, respectively. It is seen that the spin angular momentum (SAM) direction rotates anti-clockwise (clockwise) along the outer (inner) Fermi surface. The total spin polarization $|{\bf \it S}|$ = $\sqrt{{S_x}^2 + {S_y}^2 + {S_z}^2}$ and the longitudinal spin component $S_z$ along the outer and inner Fermi surfaces are given in Fig. 3(c) and 3(d), respectively. We find that the value of $|{\bf \it S}|$ in the H$_3$ structure is slightly larger than that in the T$_4$ structure. It is also found that the magnitude of the transverse spin component $S_x$ or $S_y$ is larger in the H$_3$ structure compared to the T$_4$ structure, while that of $S_z$ is reversed between the two structures. These different features between the two structures may be caused by a relatively larger in-plane potential gradient in the T$_4$ structure due to its broken $C_{3v}$ symmetry. As shown in Figs. 3(c) and 3(d), the spin vectors are highly polarized along the $z$ direction, reflecting that the Pb overlayer on the Si(111) substrate has a large in-plane component of potential gradients. It is noted that the H$_3$ structure has the $C_{3v}$ symmetry while the T$_4$ structure has one mirror-plane ${\sigma}_v$ symmetry with the $xz$ plane [see Fig. 1(b)]. Consequently, the whole spin textures along the Fermi surface satisfy the symmetries involved in the H$_3$ and T$_4$ structures, respectively [see Figs. 3(a) and 3(b)]. Since both the H$_3$ and T$_4$ structures have a mirror symmetry of the $xz$ plane, the $S_x$ and $S_z$ components change their sign but the $S_y$ component remains unchanged when the spin vectors within the irreducible part of SBZ are reflected through the mirror plane along the Fermi surface. However, the spin vectors at the points crossing the $\overline{{\Gamma}K}$ line are oriented perpendicular to the mirror plane. Note that this mirror-plane symmetry is a combination of the proper rotation of 180$^\circ$ (about the $y$ axis) with the inversion. Meanwhile, the whole spin textures shown in Figs. 3(a) and 3(b) also satisfy the time-reversal symmetry that reverses simultaneously the wavevector and spin.

\begin{figure}[ht]
\centering{ \includegraphics[width=8.0cm]{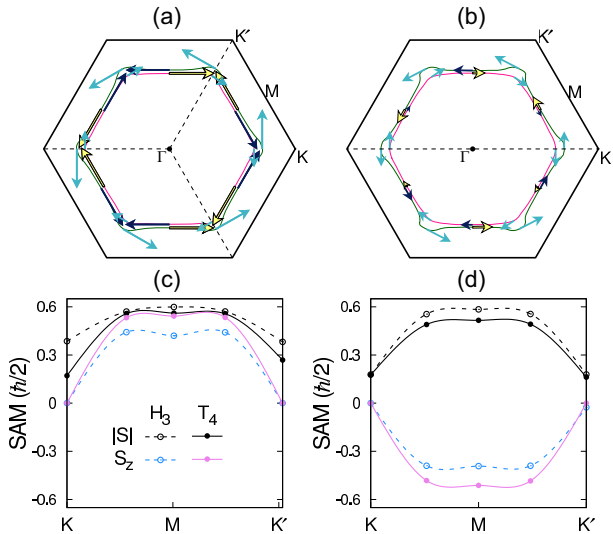} }
\caption{(Color online) Helical spin textures of the (a) H$_3$ and (b) T$_4$ structures of Pb/Si(111) along the Fermi surface. The SAM vectors with the negative, zero, and positive $S_z$ components are represented with increasing brightness of the arrows. The total spin polarization $|{\bf \it S}|$ = $\sqrt{{S_x}^2 + {S_y}^2 + {S_z}^2}$ and the $S_z$ component along the outer and inner Fermi surfaces are given in (c) and (d), respectively.}
\end{figure}

To find the energy barrier for the phase transition between the H$_3$ and T$_4$ structures, we calculate the energy profile along the transition path by using the nudged elastic-band method~\cite{neb}. The calculated energy profile for the transition state ($TS$) is displayed in Fig. 4. We find that $TS$ is higher in energy than the H$_3$ structure, yielding an energy barrier of ${\sim}$0.59 (0.27) eV on going from the H$_3$ to the T$_4$ structure in Pb/Si(111) [Pb/Ge(111)]. The presence of a sizable energy barrier between the H$_3$ and T$_4$ structures suggests that the two energetically competing structures can coexist in Pb/Si(111) or Pb/Ge(111) at low temperatures. However, as temperature increases, thermal fluctuation between the two structures is plausible. Based on an Arrhenius-type activation process with the usual attempt frequency of ${\sim}$10$^{13}$ Hz, we estimate the thermal fluctuation temperature as ${\sim}$270 and ${\sim}$125 K for Pb/Si(111) and Pb/Ge(111), respectively. This peculiar feature of the two competing structures separated by a sizable energy barrier is likely to cause not only the existence of complicated mixed phases at low temperatures but also the order-disorder transition at high temperatures, as observed in the Pb/Si(111) system.~\cite{hori,step}

\begin{figure}[ht]
\centering{ \includegraphics[width=8.0cm]{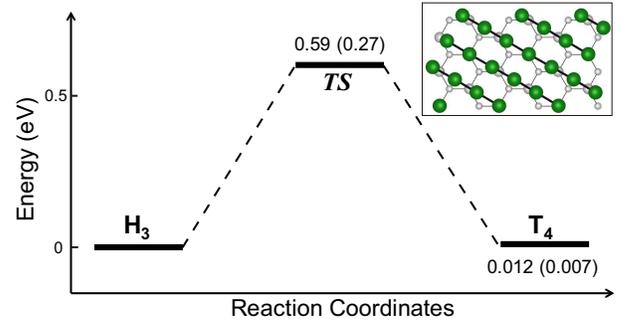} }
\caption{(Color online) Calculated energy profile along the transition pathway from the H$_3$ to the T$_4$ structure in Pb/Si(111). The atomic geometry of the transition state (TS) is given. The numbers denote the total energies of TS and the T$_4$ structure relative to the H$_3$ structure. The corresponding energy values for Pb/Ge(111) are given in parentheses.}
\end{figure}

\section{SUMMARY}

We have investigated the 4/3-ML Pb overlayer structures on the Si(111) and Ge(111) surfaces within the H$_3$ and T$_4$ models using the DFT calculations with/without including the SOC. For both Pb/Si(111) and Pb/Ge(111) systems, the DFT calculation without SOC showed that the T$_4$ structure is energetically favored over the H$_3$ structure, but the DFT+SOC calculation reverses their relative stability. This SOC-driven switching of the ground state is accounted for in terms of a more asymmetric surface charge distribution in the H$_3$ structure compared to the T$_4$ structure, which in turn gives rise to a relatively larger Rashba spin splitting as well as a relatively larger pseudo-gap opening along the $\overline{KM}$ line. In addition, our nudged elastic band calculations for Pb/Si(111) and Pb/Ge(111) showed the presence of a sizable energy barrier between the H$_3$ and T$_4$ structures, so that the two energetically competing structures can coexist at low temperatures. Further experiments are needed for identifying such a close proximity of thermodynamically phase-separated ground states.

\vspace{0.4cm}
\noindent{\bf ACKNOWLEDGEMENTS }
\vspace{0.4cm}

This work was supported by National Research Foundation of Korea (NRF) grant funded by the Korea Government (MSIP) (NRF-2014M2B2A9032247) and by the National Basic Research Program of China (No.2012CB921300), the National Natural Science Foundation of China (No.11274280), and Innovation Scientists and Technicians Troop Construction Projects of Henan Province. The calculations were performed by KISTI supercomputing center through the strategic support program (KSC-2015-C3-044) for the supercomputing application research.

\noindent $^{*}$ Corresponding authors: chojh@hanyang.ac.kr and jiayu@zzu.edu.cn


\end{document}